# BERT_SE: A Pre-trained Language Representation Model for Software Engineering

Eliane Maria De Bortoli Fávero and Dalcimar Casanova

Department of Informatics, Technological University of Paraná, Curitiba, Brazil

## Abstract

*The application of Natural Language Processing (NLP) has achieved a high level of relevance in several areas. In the field of software engineering (SE), NLP applications are based on the classification of similar texts (e.g. software requirements), applied in tasks of estimating software effort, selection of human resources, etc. Classifying software requirements has been a complex task, considering the informality and complexity inherent in the texts produced during the software development process. The pre-trained embedding models are shown as a viable alternative when considering the low volume of textual data labeled in the area of software engineering, as well as the lack of quality of these data. Although there is much research around the application of word embedding in several areas, to date, there is no knowledge of studies that have explored its application in the creation of a specific model for the domain of the SE area. Thus, this article presents the proposal for a contextualized embedding model, called BERT_SE, which allows the recognition of specific and relevant terms in the context of SE. The assessment of BERT_SE was performed using the software requirements classification task, demonstrating that this model has an average improvement rate of 13% concerning the BERT_base model, made available by the authors of BERT. The code and pre-trained models are available at https://github.com/elianedb.*

## Keywords

*Word embedding, Software engineering, Domain-specific Model, Contextualized pre-trained model, BERT*

## 1. Introduction

The software development process requires some indispensable activities in its planning phase, such as effort estimates for the project requirements, the selection of adequate human resources for development, the search for reusable resources, among others. Often, the information needed to search for such resources is in text format (e.g. use cases, user stories, bug reports). To obtain this information is a very complex task, considering the intrinsic informality in many software development processes regarding the textual artifacts that are produced in them. This difficulty occurs mainly because, in addition to not having a standard structure, these texts include a diversity of domain-specific information, such as source code, links, IP addresses, among others.

A very common aspect in these texts is the occurrence of different words, but they are used in the same context, in which case there is the need to consider them similar because, although the texts are different, their context is similar. In this case, a context can be defined as the text that precedes and/or follows a particular word, sentence, or text, and that contributes to its interpretation and meaning [1]. The context is directly related to the semantics of a word concerning the situation in which it is applied. Therefore, the need to recognize domain specific terms and their meaning in each situation in which are applied is highlighted in SE, aiming to obtain the most accurate feedback possible in carrying out various tasks of area.

In this paper, the specific-domain terms are a set of common words in a particular area (e.g. medicine, software engineering), among them, there are strong semantic relations. Some studies have already been made to obtain similar words in SE area [2], similar to WordNet, which consists basically in a thesaurus, grouping words based on their meanings [3], which remain static until they are updated. An example of the need for context representation in SE would be for the distinction of ambiguous words, like in the sentence: "The implementation depends on the language". In this case, it is necessary to consider the target word (language) as its context, to infer that this is a programming language and not a speaking language.

Considering the nature of texts in SE, the application of text representation methods based on the characteristics of each word (e.g. bag-of-words) has several limitations. These limitations interfere in the generation of a learning model and are caused by: sparsity, high dimensionality, and as a result, overfitting. Furthermore, this characteristics type is not sufficient to discriminate against each requirement. It occurs because the software textual requirements require a deep analysis, in which, besides the individual characteristics of each word, semantic characteristics are obtained. It means to analyze the meaning of words contained in a requirement related to your context.

Actually, the word embedding models are a strong tendency in NLP. The first word embedding model that utilized artificial neural networks was published in 2013 [4] by Google researchers, and since then, this concept has been part of majority of research in NLP. The word embedding methods aim to mainly capture the semantic of a particular word in a specific context. This method allows the words to be represented in a dense way and with low dimensionality, facilitating applications of machine learning in which NLP is applied. The innovations in the train of word embedding models for a variety of purposes have begun in recent years with the emergence of Word2Vec [4] and GloVe [5], allowing models to learn from rather bulky corpus. So, the contextual representations by means of embedding models have been very useful in the identification of context-sensitive similarity [6], disambiguation of the word meaning [7], [8], induction of the word meaning [9], lexical substitution for the creation of an embedding generic model [10], sentence complementation [11], among others.

The methods to obtain embedding have been considered one of the greatest advanced, and prominent in the area of NLP in recent years. Different methods for the generation of embedding from texts have been developed [12], which is possible to classify them into [13]: context-less or static, and contextualized or dynamic. Contextual embedding has been considered a revolution in NLP. This approach produces different vector representations to the same word in a text, which varies according to its context and, therefore, they are able of capturing contextual semantics of ambiguous words [14], in addition to addressing polysemy issues. In this way, each occurrence of a word is mapped to a dense vector, considering specifically the surrounding context.

Some studies using embedding have been carried out specifically in the field of SE, such as the recommendation of specific-domain topics in Stack Overflow question tags [15], the recommendation of similar bugs [16], sentimental analysis in software engineering [17], ambiguity detection in requirements engineering [18], among others. None of them used contextualized embedding. Therefore, within this new paradigm, the Bidirectional Encoder Representations from Transformers (BERT) [19] have been one of the algorithms which presented the best results in NLP tasks, besides being open-source. With all these benefits, this model has been widely used in various NLP applications. Its use has occurred via pre-trained models and is available by the authors [19]). These models were pre-trained in a large corpus of unlabeled texts and of general-purpose (e.g. Wikipedia).

The pre-training of language models proved to be highly effective in learning language universal representations from unlabeled data on a large scale [20]. Therefore, there is no need of train from scratch, due to fine-tuning techniques [21]. According to the authors, this technique consists in tuning a generic pre-trained model, using an unlabeled data corpus in the domain of a particular application. This process allows to economize many hours of training and spare the need of the specific rather bulky corpus. Although there are many researches around the application of word embedding in several areas, until where it is known, so far there is no effective contextualized pre-trained model for the SE area. Therefore, this study suggests the generation of a contextualized pre-trained embedding model, able to recognize specific, and relevant terms of the area. Thus, it becomes possible to identify the specific semantic of each parsed sentence. In addition, the existence of this model seeks to help in pattern recognition among these texts, allowing its effective application in several machine learning tasks in the area, which are based on textual data (e.g.bug classification, software effort estimation based on analogy, and others).

Thus, first, the fine-tuning of the generic BERT embedding model, made available by its authors, was performed. Next, this same pre-trained model went through the fine-tuning process, used for that, a specific corpus of SE domain. The result of the fine-tuning process was a contextualized pre-trained model for SE (BERT_SE). Hence, comparative tests were performed between both models: generic and adjusted. The implementation of the approach was divided into three main steps: (1) collection and pre-processing of the used corpus; (2) preparation of data for pre-training, and (3) application of the pre-training method. Then, the evaluation of the results obtained was carried out focusing on identifying if the similarities among sentences of the context of SE, are better expressed by a generic BERT embedding model or a BERT embedding model adjusted (BERT_SE). The preliminary results obtained are promising, motivating the continuity of the research on this topic.

The following content of this paper is organized as follows. Section 2 presents the background containing relevant aspects related to word embedding as well as related works that use embedding models in specific domains. Section 3 presents the construction approach, describing the necessary steps for its implementation and the evaluation metrics used. Section 4 presents the experimental results obtained, followed by the discussion of statistics and model performance. Section 5 presents the threats to validity, followed by the conclusion and future work (session 6).

## 2. BACKGROUND

### 2.1. Pre-trained Embedding Models

The pre-training of the language model proved to be highly effective in learning universal representations of language from unlabeled data on a large scale [20]. Among the main benefits of pre-trained language models, is the fact that there is no need to train them from scratch. This characteristic, besides reducing considerably the need for computational cost, saving a lot of hours of training, become unnecessary a highly representative corpus. This is possible through fine-tuning techniques. The fine-tuning approach, also named transfer learning in some contexts, consists in introduce minor parameters of a specific task and train it in the following tasks, simply adjusting all the pre-trained parameters [19] in a generic rather bulky corpus.

Thus, pre-training has been widely applied in various NLP tasks, bringing many benefits and great advances, especially in tasks that have limited data for training [22], [21]. There are two main methods applied in the generation of pre-trained embedding: context-less and contextualized.

Word2Vec, as well as other similar models (e.g. Glove [5]) are considered algorithms for textual representations context-less. This means that these models present restrictions regarding the representation of the context of words in a text, impairing tasks at the sentence level or even fine-tuning at the word level. This is because these models are unidirectional, that is, they consider the context of a word only from left to right, with no mechanism that detects if a particular word has already occurred in the corpus before. Therefore, these models provide a single representation, using a dense vector, for each word in a text or set of texts.

In addition, according to its authors [4] these models are considered very shallow, as they represent each word in only one layer, and there is a limit on the amount of information they can capture. Finally, these models do not consider the polysemy of words, that is, the same word being used in different contexts can have different meanings (e.g. bank – monetary sense; bank - to sit), which is not treated by these models. Another characteristic that is not treated is the ambiguity of the words, that is, when two or more different words have the same meaning (e.g. create, implement, generate).

On the other hand, many advances have occurred in the area of NLP in recent years. Such advances are due, mainly, to deep learning techniques [23]. Among these advances is the possibility of obtaining contextualized embedding. This approach produces different vector representations for the same word in a text, which varies according to its context. Therefore, these techniques are capable of capturing contextual semantics of ambiguous words [14], as well as addressing polysemy issues. From this new paradigm, recent studies have turned to research that applies contextualized embedding models [24] [14], leaving aside the original paradigm, in which there was only one vector of embedding for each single word in one text/set of texts. Thus, each occurrence of a word is mapped to a dense vector, specifically considering the surrounding context.

This representation approach is easily applicable to many NLP tasks, where the inputs are usually sentences and therefore, the context information is available, such as textual software requirements. This new language representation paradigm originated from several ideas and initiatives that emerged in NLP in recent years, such as: coVe [25], ELMo [24], ULMFiT [21], CVT [26], Context2Vec [10], BERT [19] and Transformer OpenAI (GPT e GPT-2) [27]. The BERT contextualized pre-trained model [19], has presented results greatly improved in NLP tasks, and has therefore been widely used in several applications. Its application has occurred through pre-trained models and available by its authors (e.g. BERT_base e BERT large [19]).

## 2.2. BERT

The BERT is an innovative method, considered the state of the art in pre-trained language representation [19]. BERT models are considered contextualized or dynamic models, and have shown much-improved results in several NLP tasks [22], [24], [27], [21] as sentiment classification, calculation of semantic tasks of textual similarity and recognition of tasks of textual linking.

This model originated from various ideas and initiatives aimed at textual representation that have emerged in the area of NLP in recent years, such as: coVe [25], ELMo [24], ULMFiT [21], CVT [26], context2Vec [28], the OpenAI transformer (GPT and GPT-2) [27] and the Transformer [29]. BERT is characterized as a dynamic method, mainly because it has an attention mechanism, also called Transformer [19], which allows analyzing the context of each word in a text individually, including checking if each word has been previously used in a text with the same context. This allows the method to learn contextual relationships between words (or subwords) in a text. BERT consists of several Transformer models [29] whose parameters are pre-trained on an unlabeled corpus like Wikipedia and BooksCorpus [30]. It can say that for a given input sentence, BERT "looks left and right several times" and outputs a dense vector representation for each word. For this reason, BERT is classified as a profoundly two-way model because it learns two representations of each word, one on the right and one on the left, and this learning to repeat n times. These representations are concatenated to obtain a final representation to use in future tasks.

The pre-processing model adopted by BERT accomplishes two main tasks: masked language modeling (MLM) and next sentence prediction (NSP). In the MLM task, the authors argue [19] that it is possible to predict a particular masked word from the context. For example, let's say we have a phrase: "I love reading data science articles." We want to train a contextualized language model. In this case, you need to replace "data" with "[MASK]". It is a token to indicate that it is missing. We will then train the model so that it can predict "date" as the missing token: "I love reading articles from [MASK] science".

This technique aims to make the model learn the relationship between words, improving the level of learning, avoiding a possible "vicious cycle", in which the prediction of a word to base on the word itself. Devlin et al. [19] used 15-20% of words as masked words.

The task of NSP is to learn the relationship between sentences. As with MLM, given two sentences (A and B), we want to know if B is the next sentence after A in the corpus or if it would be any sentence.

With this, BERT combines the pre-training tasks of both tasks (MLM and NSP), making it a task-independent model. For this, their authors provided pre-trained models in a generic corpus but allowing fine-tuning. It means that instead of taking days to pre-workout, it only takes a few hours. According to the authors of BERT [19], a new state of the art has been achieved in all NLP tasks they have attempted (e.g. Question Answering (QA) and Natural Language Inference (NLI)).

## 2.3. Work-related to the use of specific domains

In recent years, several initiatives making use of textual representations have been applied in several domains. But some areas require representations of words that consider particularities of a particular domain (e.g. health, technology, software engineering). The following will present some of the studies that involve the representation of textual data from specific domains.

A study by [31] addresses the task of extracting events from a representation model of a domain-specific dataset. Is described a set of participants (i.e. attributes or roles) whose values are text excerpts. The authors show that learning word representations from unlabeled domain-specific data and using them to represent event roles enable them to outperform previous state-of-the-art event, extraction models.

Another application occurred in biomedical text mining. In this area, there are many entities and syntactic parts that present rich domain information. In this way, [32], presented a model of word embedding specific to this domain.

Focusing on the ES area will be presented in the sequence some research that has explored aspects of specific domains. The approach of a recommendation system of similar libraries of software implementation (e.g. similar to JUnit) was proposed by [15]. This approach solves queries about these libraries by combining the word embedding technique and domain-specific knowledge extracted from millions of Stack Overflow tags. Still in this line, [2] proposed SEWordSim. It is a lexicon, that is, a dataset of similar words specific to the SE domain. The similarity characteristics between the words were extracted automatically from the questions and answers available in Stack Overflow.

Another study proposed by Celefato et al. [17] addresses the problem of applying feelings analysis to the discipline of software engineering. This classifier uses a set of semantic resources based on a domain-dependent lexicon. Ye et al. [33] explored word embedding to improve information retrieval in software engineering. Its ultimate goal is to eliminate the lexical gap between code fragments and natural language descriptions that can be found in tutorials, API documentation, and bug reports. They empirically demonstrate how exploring word embedding improves next-generation approaches to bug tracking.

Word embedding techniques were also applied to estimate the degree of ambiguity of words typical of the context of computer science (e.g. system, database, interface) when used in different application domains [18]. The results show that it is possible to identify variations of the meaning of the terms of computer science in the applied domains, providing an estimate of the distance between the considered domains. A new approach to recommending similar bugs was proposed [16]. The approach combines standard information retrieval techniques and word embedding techniques. Sugathadasa et al. [34] proposed new measures of semantic similarity aimed at specific domains. This measure was created by the synergetic union of word2vec and lexical-based semantic similarity methods.

There is a wide variety of studies that use embedding models, but our proposal differs from these approaches because, firstly, it aims to overcome the limitations imposed by bag-of-words models, especially concerning dimensionality and sparsity. Also, the differential of the proposed study is in its application. Because it offers the pre-training word embedding model and allows a multitasking representation of textual artifacts to perform tasks involving NLP in the SE domain.

According to Chen et al. [15], multi-task learning requires that tasks be trained from scratch at a time, which makes it inefficient and often requires careful consideration of the task's specific objective functions. Thus, one of the great advantages of this model is that there is no need to train a model from scratch, or even have a massive corpus representing words and their semantic relationships within that domain.

## 2. PRE-TRAINED MODEL FINE-TUNING PROCESS

The main objective of this article is to present BERT_SE, a pre-trained language representation model and adjusted for the SE domain. To do this, we first started BERT_SE with the standard weights of BERT_base [19], which was pre-trained in a general domain corpus (Wikipedia in English and Books Corpus).

As a generic model of BERT, the *BERT_base uncased* was applied, which was previously trained and became available by its authors [19] for free use in NLP tasks. Table 1 presents the specifications for this model.

Table 1. BERT pré-trained model used in the proposed approach.

| Pre-trained model | Specification |
| --- | --- |
| BERT base | BERT_base uncased: 12 layers for each token, 768 hidden layers, 12 heads of attention, 110 million parameters. The uncased specification means that the text was converted to lower case before tokenization based on WordPiece, in addition, removes any accent marks. This model was trained with English texts (Wikipedia) with lowercase letters. |

The BERT_base model, as well as the other pre-trained BERT models, offers 3 components [19]:

- A TensorFlow checkpoint (*bert_model.ckpt*) that contains pre-trained weights (consisting of 3 files).

- A vocabulary file (*vocab.txt*) for mapping WordPiece [35] for word identification.

- A configuration file (*bert_config.json*) that specifies the model's hyperparameters.

## 2.1. Datasets Used for the Fine-tuning

The BERT_base fine-tuning process is performed using a corpus of the SE domain, here called corp_SE. The composition of the corp_SE is shown in Table 2.

A Stack Overflow dataset was chosen because it is relatively restricted, contains a large number of posts and associated tags, and the data is easy to obtain. It is restricted because it refers to a specific domain (of software engineering), that the authors (users of the forum) can use, which is different from a less restricted domain (e.g. Twitter). The remaining corpus used is all derived from open-source software projects or from software development companies that have authorized their application in research in the area. These data were in a .csv file. After that, a basic pre-processing was carried out, with the objective of excluding special characters, HTML tags, and numbers. Thus, in order to maintain a standard, the same pre-processing applied to software requirements data, obtained from open source projects and used by Choetkiertikul et al. (2018), was performed. For the pre-processing, a method based on specific regular expressions was applied. Table 3 presents some examples of sentences that composed the corp_SE.

Therefore, the corp_SE is composed of 456.500 texts, in this paper called sentences. Each sentence has an average length of 61 words. The vocabulary generated by the corp_SE is composed of 1.179.501 words.

## 2.2. Performing of Fine-tuning

It stands out that the fine-tuning process consists of the use of a pre-trained embedding model from a generic dataset in an unsupervised way, which is adjusted, that is, retrained on a known dataset that is specific to the area of interest. In this case, the fine-tuning was performed on the generic model BERT_base, using corp_SE (according to Table 2). The fine-tuning process of the pre-trained BERT model consists of two main steps [19]:

1) *Preparation of data for pre-training:* initially, the input data is generated for pre-training. This is made by converting the input sentences into the format expected by the BERT model (using *create_pretraining_data* algorithm). As BERT can receive one or two sentences as input, the model expects an input format in which special tokens mark the beginning and end of each sentence, as shown in Table 4. In addition, the tokenization process needs to be performed. BERT provides its own tokenizer, which generates output as shown in Table 5.

Table 2. Dataset used in the composing of corp_SE.

| Data source | Specification |
| --- | --- |
| Sentence subset of Stack Overflow (www.stackoverflow.com) | A subset of Stackoverflow sentences taken from the Kaggle [36] repository, totaling 137,474 sentences after pre-processing. These sentences consist of posts made by users about doubts and problems related to the most diverse software development technologies. |
| Software requirements (user stories) obtained from open-source projects | Textual corpus of 319.026 requirements from 16 large open-source projects in 9 repositories (Apache, Appcelerator, DuraSpace, Atlassian, Moodle, Lsstcorp, Mulesoft, Spring, and Talendforge) [37] and from others 22 open-source datasets [38]. |

|  | According to the authors that available the datasets, all were obtained online or from software companies with permission for dissemination. |
|---|---|

Table 3. Example Sentences that Composed the corp_SE (before pre-processing).

| Text ID | Text |
|---------|------|
| 1 | Create project references property pagehtml create property page for project which allows manipulation of embedded references. user can add or remove references from the filesystem or url (if possible). |
| 2 | android: permissions failure in android.calendar drillbit testlooks like the android.calendar test recently started failing due to missing permissions, log:code permission denial: opening provider com.android.providers.calendar.calendarprovider2 from processrecord(423b2048 20514:org.appcelerator.titanium.testharness 10082) (pid=20514, uid=10082) requires android.permission.read calendar or android.permission.write calendar code |
| 3 | ws security signature support for ws consumer: we should support ws security signature and verification capabilities in ws consumer. |

2) *Application of the pre-training method:* the method used for pre-training by BERT (run pretraining) became available by its authors. The necessary hyperparameters were informed, the most important being:

- *max_seq_length:* defining the maximum size of the input texts (set at 100).

- *batch-size:* maximum lot size (set at 32, per use guidance of the pre-trained model BERT base).

- *epochs number:* the standard epochs number of model is 100. This number has even been varied to 500 and 1000 during the experiments.

Table 4. Example of formatting input texts for pre-training with BERT.

| Entry of two sentences | Entry of a sentence |
|------------------------|---------------------|
| [CLS] The man went to the store. [SEP] He bought a gallon of milk.[SEP] | [CLS] The man went to the store.[SEP] |

Table 5. Example application of tokenizer provided by BERT.

| Input sentence | "Here is the sentence I want embedding for." |
|----------------|-----------------------------------------------|
| **Text after *tokenizer*** | ['[CLS]', 'here', 'is', 'the', 'sentence', 'i', 'want', 'em', '##bed', '##ding', '##s', 'for', '.', '[SEP]'] |

Justified that in transfer learning is not suggested change in the values of the hyperparameters, except for the *epochs number* of training. The *max_seq_length* was not varied, as it was observed from the exploratory analysis of the data, that a value equal to 100 would correspond to most sentences in the dataset. For *batch-size* we chose to leave the default value of 32, considering the amount of memory available. The study by Devlin et al. (2019) however, demonstrates that changes in this hyperparameter usually do not lead to large differences in performance. For the generic BERT model, we opted for its version *Uncased L-12* base, here called BERT_base (Table 1). The fine-tuning process for the BERT model was performed as shown the Figure 1.

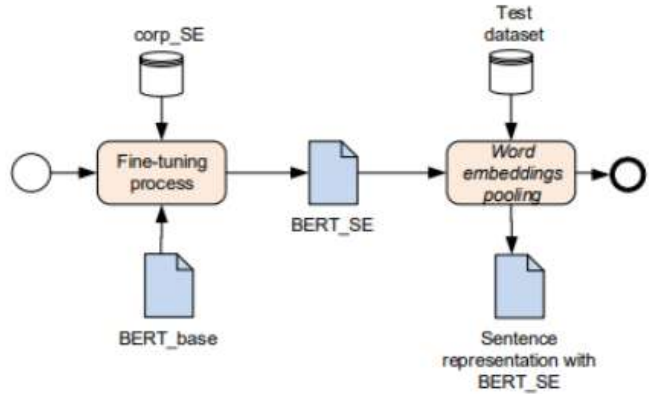

Figure 1.  Pipeline of the fine-tuning process of the BERT_base model and generation of the textual representation for the corp_SE.

The entire process, from data preparation to fine-tuning the BERT model, used the algorithms available in the repository *https://github.com/google-research/bert*, in which the authors [19] provide the full framework developed in the Python language.

After performing the fine-tuning, a new pre-trained model is available, as shown in Table 6, which will compose the experiments. It is noteworthy that the proposed model requires pre-training only for the embedding layer. This allows, for example, this pre-trained model is available for other software engineering tasks, or even for different effort estimation tasks. Thus, this pre-trained model may undergo successive adjustments, according to the need of the task to which it will be applied.

## 2.2. Performing of Fine-tuning

This step consists of analyzing the similarity between a source sentence and a target sentence. The use of the cosine similarity measure was defined considering that it is a predominant way of estimating the similarity of two documents based on word incorporation. Thus, the cosine similarity measure must be applied to the two centroids obtained from the embedding vectors associated with the words in each document [39].

That is, given a sentence set regarding to SE, was observed the cosine distance of each sentence concerning the others.

The similarity among sentences is given by the cosine distance among mean embedding vectors that represent them. This is made as to the generic model (BERT_base), as the adjusted model (BERT_SE) obtained from the trained model. That is:

$$t_i = \frac{1}{n} \sum_{i=1}^{n} p_i$$

where p = {$p_1$, $p_2$, ..., $p_n$}  is a word of the sentence t = {$t_1$, $t_2$, ..., $t_n$} with *n* elements in a vector. So, the cosine similarity of two sentences is given by:

$$cos(\mathbf{t}, \mathbf{t}') = \frac{\mathbf{t} \times \mathbf{t}'}{||\mathbf{t}|| \times ||\mathbf{t}'||}$$

where both *t* and *t'* are the mean embedding of sentences.

The cosine distance returns a value between [0; 1]. Values closer to 1 indicate greater similarity among vectors. Specialists in the field evaluated, case by case, whether the sentence of origin had semantic similarity or not. For this, a sample of 30 professionals in the field, working in different companies in the field, was selected (e.g. analysts, developers, project managers).

## 3. RESULTS AND DISCUSSIONS

The results presented in this section illustrate that a general-purpose embedding model, even though it presents a larger vocabulary, does not necessarily adequately represent area specific terms, such as SE. The following experiments intended to demonstrate that results of textual classification tasks in the SE domain could be improved if the language model used for its representation is fined using a specific context dataset.

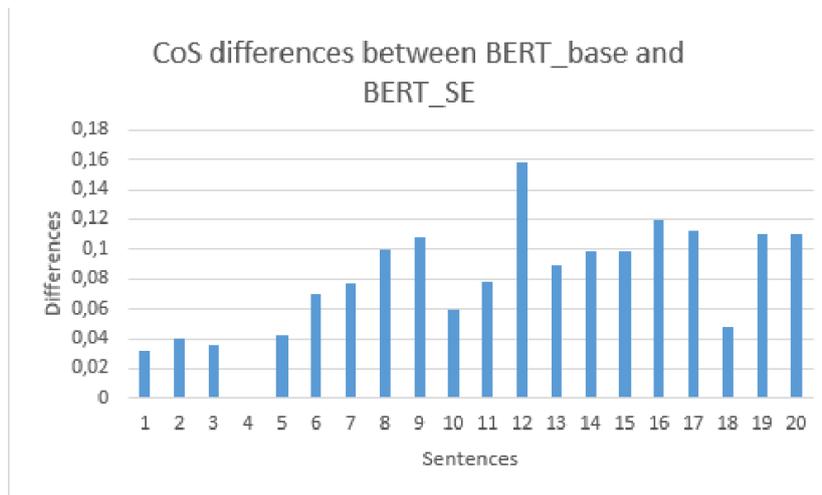

Figure 2. CoS differences between BERT_base and BERT_SE, when the sample S5 (Table 6) is compared with the remaining test sentences.

It's important to remember, that the cosine measure is a trigonometric function that provides a value equal to 1 if the understood angle is zero, that is, if both vectors point to the same place (identical objects). For any angle other than 0, the cosine value is less than one. If the vectors were orthogonal, the coSine would cancel out, and if they pointed in the opposite direction, its value would be -1. Thus, the value of this metric is between -1 and 1. Therefore, the closer the coS value gets to 1, the greater is the similarity between two sentences.

The coS values for BERT_SE are larger, and therefore, closer to 1 when compared to the same values for BERT_base. This indicates a greater cosine similarity between the sentences represented by BERT_SE. This fact can be verified in Table 6, where a BERT_SE improvement rate concerning a base of BERT can be verified.

Especially in software engineering, it is important to note the similarity of contexts among sentences. For example, in S4 (Table 6), both sentences deal with adding a new record (sales order and users). In this way, regardless of the existing content in a register, they are close to implementation operations. Similarly in S3, where both sentences refer to adding a button to a form. Therefore, the coS is expected to be high, since the source and target are from the same context. When presenting the results obtained in Table 6, and asked about the similarity between the sentences used in the evaluation of the model, the specialists approached indicated that there was similarity.

Table 6.  Cosine similarity (coS) between sentences in the SE domain, obtained from
BERT_base e BERT_SE.

| Sample ID | Source sentence | Target sentence | coS BERT_base | coS BERT_SE | BERT_SE improvement rate (%) |
|---|---|---|---|---|---|
| S1 | Create user registration allowing to Include a photo and digital | Create a button that allows you to retrieve the last record deleted from the order | 0.59 | 0.71 | 20.33 |
| S2 | Create a method that allows the user to customize sales reports | List all store products by category | 0.71 | 0.84 | 15.4 |
| S3 | Include calculation button by product in the sales order | Create a button that allows you to retrieve the last record deleted from the order | 0.65 | 0.80 | 18.75 |
| S4 | As a salesperson, I want to include sales orders | Create user registration allowing to include a photo and digital | 0.58 | 0.70 | 20.7 |
| S5 | Add user authentication function for accessing the system | As an administrator, I need to have access to a sales report to find out how much I received in a given period | 0.75 | 0.84 | 10.12 |

Figure 2 shows an experiment where the difference of coS similarities of both BERT_SE and BERT_base models is measured. If the difference is a positive value then BERT_SE has a high similarity value. If the difference has a negative value then BERT_base shows a better similarity representation. In this experiment, we measure the similarity of sentence S5 (see Table 6) against all other sentences. It is observed that all similarities increases when using BERT_SE model, except for the fourth sentence, where BERT_base and BERT_SE produced equal values.

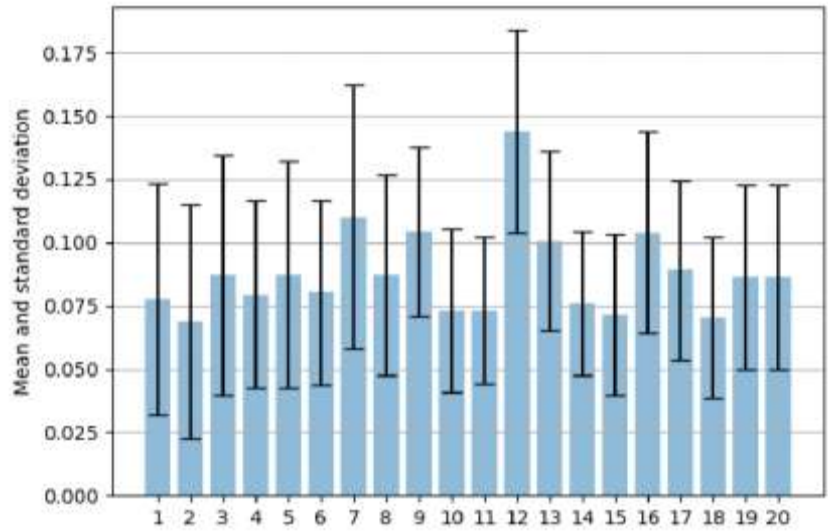

Figure 3. Mean and standard deviation for CoS differences between BERT_base and BERT_SE, for all twenty test sentences.

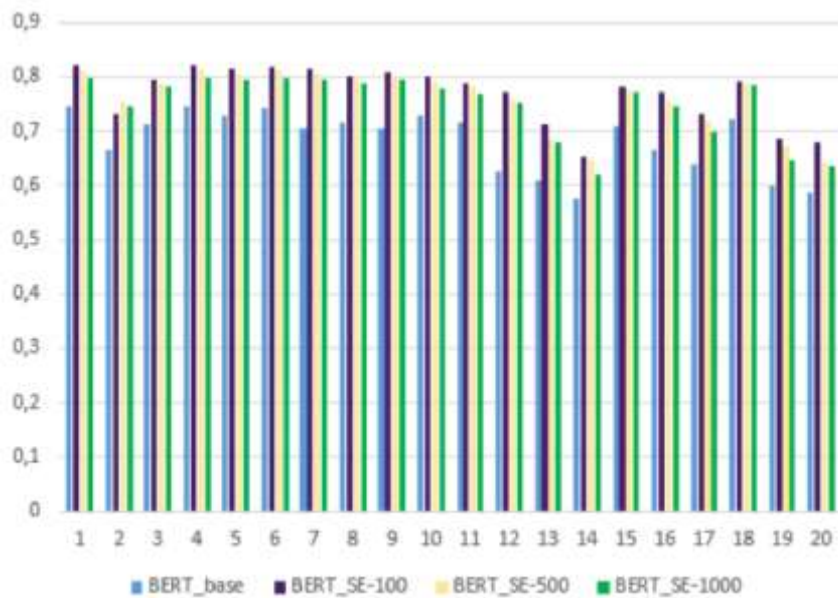

Figure 4. CoS differences mean obtained from BERT_base and BERT_SE (adjusted with different epochs number), for all twenty sentences.

An experiment similar to the previous one is shown in Figure 3. In this figure, the mean and standard deviation of the differences between the representation given by BERT_base and by BERT_SE (trained with 100 epochs), for each sentence is shown. It is observed that the average values of the differences are positive in all sentences, favoring the representation given by BERT_SE. The standard deviation, when compared to this gain, is low, which confirms this positive difference.

This result confirms that the BERT_SE increases the coS similarity if compared with BERT_base. In the next experiment, we investigate if more training time can reach even better results.

Figure 4 shows the average of the cosine distances for each sentence. First when using the BERT_base model and then when applying the adjusted model BERT_SE, trained with a different epochs number (100, 500, 1000). It is observed that the increase in the epochs number of training did not generate significant improvements in the results, but it was evident that the results are always better when applying BERT_SE.

Figure 5 shows the percentage of improvement obtained about the representation of sentences by the BERT_base model when compared with the cosine similarity values obtained when applying the BERT_SE model. It is observed that there was an improvement in all representations of sentences that used the BERT_SE model. The average improvement rate is 13% compared to the initial representation by BERT_base.

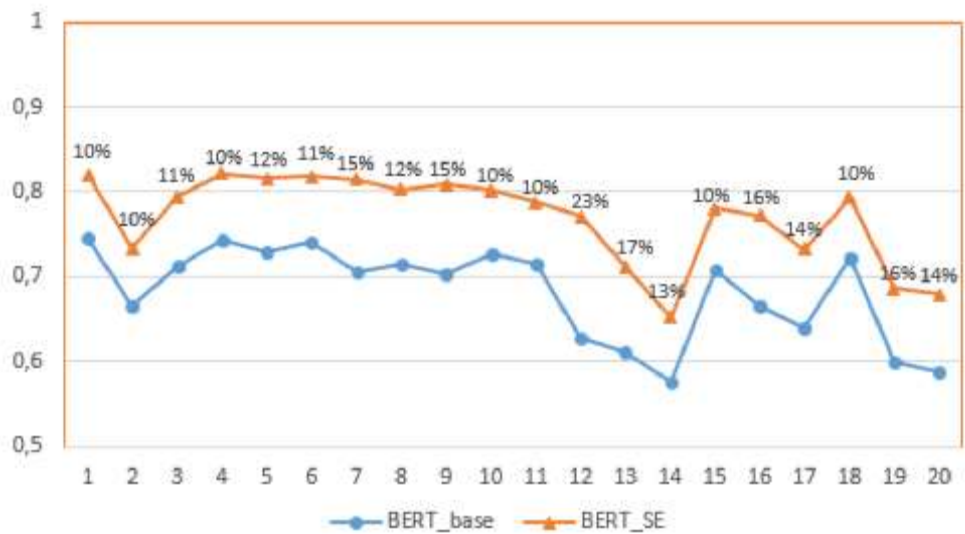

Figure 5.  Rate of improvement obtained when representing each sentence with the BERT_SE model in relation to the representation as BERT_base.

## 4. THREATS TO VALIDITY

This paper proposes the BERT SE, a contextualized pre-trained model to the textual representation in the SE domain. The results of the proposed model were positive, by testing the efficiency in the identification of similar sentences in SE. The pre-trained embedding model BERT_base, which became available by its authors [19], was generated from a dataset extremely wide, containing texts from diverse areas (e.g.Wikipedia), which makes it rather generic. Therefore, it is important to consider the overall model, which is related directly to the availability and diversity of domain data.

To generate a pre-trained embedding model and adjusted it for a specific domain, such as BERT_SE, it is necessary to have a dataset containing only texts related to the SE area, which must be in the same language and pass through the same pre-processing. For the case of SE, this data must reflect the reality of software projects in different areas, different models of the development process, forms of representation of user requirements, technologies, among other attributes. Therefore, we believe that the samples may not be sufficient to represent all the textual variations that exist in SE.

So it is recommended to update BERT_SE whenever new textual data is obtained that is appropriate. Therefore, SE is an area in constant evolution, in which new technologies often appear, another reason to keep the model updated periodically.

The tests performed with the BERT_SE model were validated by a specialists sample in the field, made up of developers, analysts, and project managers. Thus, it is necessary to consider the subjectivity intrinsic to this evaluation, which was carried out according to the opinion and experience of each one of them, which can generate a bias in the results obtained.

## 4. CONCLUSIONS

The use of pre-trained models for specific domains has shown good results in their applications, such as SciBERT [40] - a pre-trained Language Model for Scientific Text, and BioBERT [41] - a pre-trained biomedical language representation model for biomedical text mining. Therefore, this article aims to explore a BERT_SE, a contextualized pre-trained language model based on BERT, which is destined for textual classification in the field of software engineering.

The BERT_SE was generated from fine-tuning of the BERT_base model [19], a pre-trained embedding model from the generic dataset in an unsupervised way. Thus, BERT_SE was retrained in an unlabeled and specific dataset for the SE area. In this case, fine-tuning was performed on the generic BERT_base model, using the corpus corp_SE. This fine-tuning process to generate BERT_SE, confirmed the statement by the authors of BERT [19] that fine-tuning takes only a few hours on a GPU and does not require a very large specific corpus.

Thus, when compared to the pre-training process, fine-tuning is relatively inexpensive. The results are verified in a sentence representation experiment over software requirements. In this experiment, we expect that a sentence needs be more similar to each other if it belongs to the same context. The results show that the BERT_SE model surpasses the generic model in all representation tests performed, which results in an average improvement rate of 13% about initial representation, given by BERT_base.

As the results of the article show, BERT_SE behaves very well in classifying sentences in the SE area, even considering their context. Thus, it would be possible to perform several NLP tasks in this area with greater precision (e.g. bug classification, software effort estimation based on analogy, and others). The work by Fávero et al. (2020) [42] – in review by the international journal of the area - presents an application scenario for the software effort estimation analogy-based using BERT_SE, where the fine-tuning was performed with a specific less bulky dataset, and the results were positive.

As future work, we intend to launch a version of BERT_SE similar to BERT_large [19]. Besides, increase the volume of the corp_SE and generate a model with its vocabulary, to better correspond to the training corpus, and compare it with the original BERT model. It is also intended to evaluate the model in other NLP tasks (e.g. Named Entity Recognition (NER), Question Answering (QA), etc.) and that may apply to software engineering.